# Hartree–Fock variational bounds for ground state energy of chargeless fermions with finite magnetic moment in presence of a hard core potential: A stable ferromagnetic state

SUDHANSHU S. JHA [1] and S. D. MAHANTI [2]

[1]Department of Physics, Indian Institute of Technology Bombay, Mumbai 400076, India

[2]Department of Physics and Astronomy, Michigan State University, East Lansing, MI 48824-2320, USA

E-mail: ssjha@phy.iitb.ac.in ; mahanti@pa.msu.edu

**Abstract.** We use different types of determinantal Hartree-Fock (HF) wave functions to calculate variational bounds for the ground state energy of $N$ spin-half fermions in volume $V_0$, with mass $m$, electric charge zero, and magnetic moment $\mu$, which are interacting through long range magnetic dipole-dipole interaction. We find that at high densities when the average inter particle distance $r_0$ becomes small compared to the magnetic length $r_m \equiv 2m\mu^2/\hbar^2$, a ferromagnetic state with spheroidal occupation function $n_\uparrow(\vec{k})$, involving quadrupolar deformation, gives a lower energy compared to the variational energy for the uniform paramagnetic state. This HF variational bound to the ground state energy turns out to have a lower energy than our earlier calculation in which instead of a determinantal wavefunction we had used a positive semi-definite single particle density matrix operator $\rho^{(1)}$ whose eigenvalues $n_\uparrow(\vec{k})$, having quadrupolar deformation, were allowed to take any value from 0 to 1. This system is of course still unstable towards infinite density collapse, but we show here explicitly that a suitable short range repulsive (hard core) interaction of strength $U_0$ and range $a$ can stop this collapse. The existence of a stable high density ferromagnetic state with spheroidal occupation function is possible as long as the ratio of coupling constants $\Gamma_{cm} \equiv (U_0 a^3/\mu^2)$ is not very small compared to 1.

## 1. Introduction

Because of its great relevance to the properties of matter, the problem of the nature of the ground state of many–particle electron systems, like the electron gas, has been studied extensively for more than seven decades now. However, the nature of the ground state of quantum spin-half chargeless fermion systems, like neutrinos, with finite magnetic dipole moment has not drawn too much attention. In this paper, we will address this problem involving spin- dependent long range non central interaction between particles with no electric charge, with or without a short range repulsive central hard core interaction. We will use various types of determinantal Hartree-Fock (HF) wave functions to investigate the nature of the ground state and to calculate variational upper bounds to the ground state energy of such systems. This may not be of direct relevance to the neutrino cosmology at this stage, because of the tiny neutrino magnetic moment, but it is a very interesting problem in itself in the study of quantum many- particle systems.

In a recent work [1], we had considered the nature of the ground state of a system of $N$ chargeless fermions of spin-1/2, mass $m$ and magnetic moment $\mu$, interacting through the long range magnetic dipole-dipole interaction only. We had shown that at high densities when the interparticle distance $r_0$ becomes smaller than the magnetic length $r_m \equiv 2m\mu^2/\hbar^2$, a fully saturated ferromagnetic state with quadrupolar deformation of the single particle occupation function $n_\uparrow(\vec{k})$, with $n_\downarrow(\vec{k}) = 0,$ has lower energy than the energy of any variational uniform paramagnetic state with occupation function independent of spin quantum numbers (↑: $s_z$ = +1/2 or ↓: $s_z$ = -1/2). Here, $(4\pi/3)r_0^3 = V_0/N$ for the particles in volume $V_0$. The exchange energy varying as $1/r_0^3$ becomes attractive, and it dominates the repulsive kinetic energy contribution varying as $1/r_0^2$. Below some value of the dimensionless density parameter $r_{sm} \equiv (2.21)r_0/r_m$, the total energy becomes negative and the system collapses towards an infinite density state. We will explicitly show in this paper that this instability of the system towards the high-density collapse is stopped by a suitable short range repulsive finite hardcore interaction.

Our variational calculation [1], described above, was not done strictly with an N-particle determinantal HF wave function, but we had used instead eigen values $n_{\vec{k}\sigma}$ of a positive semi-definite single particle density matrix operator $\rho^{(1)} = \sum_{\vec{k}\sigma} n_{\vec{k}\sigma} |\vec{k}\sigma\rangle\langle\vec{k}\sigma|$, with $0 \leq n_\sigma(\vec{k}) \leq 1$, as variational parameters. However, Lieb[2] has shown long back that unless the interaction is purely repulsive everywhere, which is not true in our problem, only a variational determinantal HF wavefunction can give an upper bound to the ground state energy, and not any arbitrary variational positive definite single particle density matrix. In Sec 2 of this paper, we will discuss this aspect in greater detail, and use different forms of determinantal HF functions to obtain the upper bound for the ground state energy of the chargeless fermions interacting through only magnetic dipole-dipole interaction. At high enough densities, a ferromagnetic state with a spheroidal shape of the boundary of occupied states in $\vec{k}\uparrow$-space ($n_{\vec{k}\uparrow} = 1$), to be referred as the JM ferromagnetic state to distinguish it from the state used in [1], gives the upper bound to the ground state energy, at least for small deformations. This state, as expected, is unstable towards a high density collapse, as was the case in our earlier calculation [1].

In Sec. 3, we calculate the effect of including a short range repulsive hard core interaction of strength $U_0$ and range $a$ between the chargeless fermions, in addition to the magnetic dipole interaction. The calculation is again done using the framework of the variational determinantal HF procedure of the previous section. We find that if the hard core coupling constant $U_0 a^3$ is not extremely weak compared to the magnetic coupling constant $\mu^2$, there is no infinite density collapse of the JM ferromagnetic state. We thus show the possible existence of a stable equilibrium high density JM ferromagnetic state for such a system.

**2. Nature of variational bounds and single determinant HF variational calculations in the absence of any hard core interaction**

An N-particle determinantal HF wave function can be written in the form

$$\Psi_N(\vec{r}_1 s_1,......,\vec{r}_N s_N) = \langle \vec{r}_1 s_1,.....,\vec{r}_N s_N | \Psi_N \rangle = (1/\sqrt{N!}) \det[f_{\vec{k}_i \sigma_i}(\vec{r}_j s_j)] \quad (1)$$

as constructed from any N single particle orthonormal functions $f_{\vec{k}_i \sigma_i}(\vec{r},s)$, where $\{\vec{r},s\}$ refer to space-spin variables. The actual choice of the complete set of single particle orbitals $f$, in general, need not correspond to plane wave states. In other words, in general the quantum number $\vec{k}$ need not refer to a wave vector. In the HF case, the N-particle density matrix operator $\rho_{HF}^{(N)} = |\Psi_N><\Psi_N|$ corresponds to a pure state with matrix elements

$$\rho_{HF}^{(N)}(\vec{r}_1 s_1,......,\vec{r}_N s_N; \vec{r}_1' s_1',......,\vec{r}_N' s_N') =$$

$$\Psi_N(\vec{r}_1 s_1,.....,\vec{r}_N s_N)\Psi_N^*(\vec{r}_1' s_1',.....,\vec{r}_N' s_N') \qquad (2)$$

The corresponding reduced single particle density matrix is obtained in the usual way [2] by integrating the expression (2) over all the particle variables except one and multiplying it by the number of particles $N$. It is given by the expression

$$<\vec{r}s|\rho_{HF}^{(1)}|\vec{r}'s'> = \rho_{HF}^{(1)}(\vec{r}s,\vec{r}'s') = \sum_{\vec{k}\sigma} n_{\vec{k}\sigma} f_{\vec{k}\sigma}(\vec{r}s) f_{\vec{k}\sigma}^*(\vec{r}'s') \qquad (3a)$$

$$\rho_{HF}^{(1)} = \sum_{\vec{k}\sigma} n_{\vec{k}\sigma} |\vec{k}\sigma><\vec{k}\sigma| \qquad (3b)$$

with

$$Tr(\rho_{HF}^{(1)}) = \sum_s \int d^3r \rho_{HF}^{(1)}(\vec{r}s,\vec{r}s) = \sum_{\vec{k}\sigma} n_{\vec{k}\sigma} = N \qquad (3c)$$

where in the particular case of plane wave single particle orbitals and spin functions $\chi_\sigma$,

$$f_{\vec{k}\sigma}(\vec{r}s) = <\vec{r}s|\vec{k}\sigma> = (V_0)^{-1/2} \{\exp(i\vec{k}\cdot\vec{r})\} \chi_\sigma(s) \qquad (3d)$$

and where $n_{\vec{k}\sigma} \equiv n_\sigma(\vec{k}) = 1$ for all the N orthonormal single particle states used in constructing the determinantal HF wavefunction, and $n_\sigma(\vec{k}) = 0$ for all the other states in the complete set of single particle states in the Hilbert space.

In our earlier calculation [1], we had used $0 \leq n_\sigma(\vec{k}) \leq 1$ itself as a variational function, corresponding to an admissible [2] positive semi-definite single particle density matrix operator

$$\rho^{(1)} = \sum_{\vec{k}\sigma} n_\sigma(\vec{k})|\vec{k}\sigma><\vec{k}\sigma| \text{ , with } Tr(\rho^{(1)}) = \sum_{\vec{k}\sigma} n_\sigma(\vec{k}) = N \quad (4)$$

$$<\phi|\rho^{(1)}|\phi> = \sum_{\vec{k}\sigma} n_\sigma(\vec{k})|<\phi|\vec{k}\sigma>|^2 \leq \sum_{\vec{k}\sigma}|<\phi|\vec{k}\sigma|^2 = <\phi|\phi> \quad (5)$$

for any possible state $|\phi>$ in the relevant Hilbert space. As in the HF case (see, eq. 3a), the form of the matrix elements of $\rho^{(1)}$ remains the same,

$$\rho^{(1)}(\vec{r}s, \vec{r}'s') = \sum_{\vec{k}\sigma} n_\sigma(\vec{k}) f_{\vec{k}\sigma}(\vec{r}s) f^*_{\vec{k}\sigma}(\vec{r}'s') \quad (6)$$

but the occupation function $n_\sigma(\vec{k})$ can take any value from 0 to 1. Of course, the more general single particle density matrix operator $\rho^{(1)}$ does not correspond to $\rho^{(1)}_{HF}$, *except* in the special case when each $n_\sigma(\vec{k})$ is restricted to the value which is either 1 or 0.

Lieb [2] has shown that one may use any admissible $\rho^{(1)}$ for the single particle density matrix operator, as, e.g., given by eqs (4) and (5), and obtain the total energy of the system variationally by calculating

$$E(\rho^{(1)}) = Tr(\rho^{(1)} H_0) + (1/2)Tr(\rho^{(2)} V) = \sum_s \int d^3 r \rho^{(1)}(\vec{r}s, \vec{r}s) H_0(\vec{r}s)$$

$$+ (1/2)\sum_{s_1}\sum_{s_2} \int d^3 r_1 \int d^3 r_2 \rho^{(2)}(\vec{r}_1 s_1, \vec{r}_2 s_2; \vec{r}_1' s_1', \vec{r}_2' s_2') V(\vec{r}_1 s_1, \vec{r}_2 s_2) \quad (7)$$

where the total Hamiltonian of the system is given by

$$H = \sum_{i=1}^{N} H_0(\vec{r}_i s_i) + \sum_{i<j}^{N}\sum_{j}^{N} V(\vec{r}_i s_i, \vec{r}_j s_j). \quad (8)$$

Here, $H_0(\vec{r}_i s_i)$ represents the single particle part of the total Hamiltonian, which in our case is just the kinetic energy $-\hbar^2 \vec{\nabla}_i^2/2m$, $V(\vec{r}_1 s_1, \vec{r}_2 s_2)$ represents the interaction between particles 1 and 2, and $\rho^{(2)}$ is the two-particle density operator with matrix elements

$$\rho^{(2)}(\vec{r}_1 s_1, \vec{r}_2 s_2; \vec{r}_1' s_1', \vec{r}_2' s_2') = \rho^{(1)}(\vec{r}_1 s_1, \vec{r}_1' s_1')\rho^{(1)}(\vec{r}_2 s_2, \vec{r}_2' s_2')$$

$$- \rho^{(1)}(\vec{r}_1 s_1, \vec{r}_2' s_2')\rho^{(1)}(\vec{r}_2 s_2, \vec{r}_1' s_1') \quad (9)$$

It should be noted that when $\rho^{(1)}$ corresponds exactly to $\rho^{(1)}_{HF}$ of eq. (3), the corresponding $\rho^{(2)}$ is nothing but the reduced two-particle density matrix operator $\rho^{(2)}_{HF}$ which is obtained from $\rho^{(N)}_{HF}$ of eq.(2), and in that special case $E(\rho^{(1)})$ of eq. (7) gives exactly the variational HF energy corresponding to the chosen determinantal wave function. Lieb showed further that whereas for any admissible $\rho^{(1)}$ one can always construct a $N$-particle density matrix operator $\tilde{\rho}^{(N)}$ satisfying the Pauli principle ( not necessarily unique) such that the corresponding single particle reduced density matrix $\tilde{\rho}^{(1)} = \rho^{(1)}$, the corresponding energy $E(\rho^{(1)})$ of eq.(7) gives the variational bound

$$E_{exact} \leq E_{HF}(\rho^{(1)}_{HF}) \leq E(\rho^{(1)}) \tag{10}$$

*only* if the interaction $V(\vec{r}_1 s_1, \vec{r}_2 s_2)$ is nonnegative for all the values of its arguments. Here, $E_{exact}$ is the exact ground state energy of the system, and $E_{HF}$ is the best (lowest possible) variational HF energy. Although, this non-negativity condition is satisfied in the familiar case of the Coulomb interaction between particles of same charge, this is not true in our case of magnetic dipole-dipole interaction. In other words, we are not aware of any general proof to show that $E(\rho^{(1)})$ calculated by us in the earlier paper [1] is an upper bound to the exact ground state energy, although it may well be true for the specific choice of $\rho^{(1)}$ of eq. (4) in which we had taken a very simple form of $n_\sigma(\vec{k})$ as,

$$n_\downarrow(\vec{k}) = 0, n_\uparrow(\vec{k}) = \{(1-\beta) + \beta \; P_2(\cos\theta_{\vec{k}})\}\Theta(k_{F\uparrow} - k), 0 < \beta \leq 2/3 \tag{11}$$

where $P_2(z)$ is the Legendre polynomial $\tfrac{1}{2}(3z^2 - 1)$ and $\Theta(x)$ is the step function.

Note that whether the values of the interaction $V(\vec{r}s, \vec{r}'s')$ are positive or negative, in all cases any determinantal HF wave function corresponding to $\rho^{(1)} = \rho^{(1)}_{HF}$ of eq. (3) *always* satisfies the variational bound

$$E_{exact} \leq E_{HF}(\rho^{(1)} = \rho^{(1)}_{HF}) \tag{12}$$

It what follows next, we therefore propose to use only suitable determinantal HF wave functions to calculate the variational ground state energy of the dipole system.

## 2.1 *Single determinant variational calculations*

Let us assume that we are dealing with a determinantal HF wave function $\Psi_N$ of the type given by eq.(1) to calculate the variational energy $<\Psi_N | H | \Psi_N>$, which is equivalent to using the single particle density matrix operator $\rho^{(1)} = \rho^{(1)}_{HF}$ of eqs (3a-3d) with $n_\sigma(\vec{k}) = 1$ (0) for the occupied(unoccupied) states. In this Section, we will only consider the case with no hard core interaction. As shown in our earlier paper [1], for the magnetic dipole-dipole interaction the total Hamiltonian is

$$H = \sum_{i=1}^{N} p_i^2 / 2m + \sum_{i<}^{N} \sum_{j}^{N} V(\vec{r}_i s_i, \vec{r}_j s_j) \tag{13a}$$

$$V(\vec{r}_1 s_1, \vec{r}_2 s_2) = (\mu^2 / r^3)[\vec{s}_1 . \vec{s}_2 - 3 \vec{s}_1 . \hat{r} \; \vec{s}_2 . \hat{r}] \tag{13b}$$

$$\vec{r} = \vec{r}_1 - \vec{r}_2 \; , r = |r| \; , \hat{r} = \vec{r}/r \tag{13c}$$

$$V_{12}(\vec{q}) \equiv \int d^3 r e^{-i\vec{q}.\vec{r}} V(\vec{r}_1 s_1, \vec{r}_2 s_2) = \mu^2 \sum_{M=-2}^{+2} h_{-M} N_{12}^{(M)}(\vec{s}_1, \vec{s}_2) Y_{2,-M}(\hat{q}) \{1 - \delta_{\vec{q},0}\} \tag{13d}$$

where $h_{-M}$ are constants, $N_{12}^{(M)}(\vec{s}_1, \vec{s}_2)$ are spin operators [1] corresponding to two spin-1/2 particles, and $Y_{2M}(\hat{q})$ are spherical harmonics of order 2. The interaction matrix element vanishes for the momentum transfer $\vec{q} = 0$ and is independent of the magnitude of the momentum transfer, depending only on its direction $\hat{q}$. The two-particle spin operator $N_{12}^{(M)}$ connects only those states for which the total z-components of the two spins differ by M. When one is taking the expectation value of $V$ in any chosen determinantal state, only M=0 term contributes, the matrix elements of which are given below explicitly. In fact, the total variational energy is then given by

$$E = <\Psi_N | H | \Psi_N> = E_{kin} + E_{exch} \tag{14}$$

$$E_{kin} = \sum_{\vec{k}\sigma} (\hbar^2 k^2 / 2m) \; n_\sigma(\vec{k}) \tag{15}$$

$$E_{exch} = -(\mu^2 / 2) h_0 \frac{1}{V_0} \sum_{\vec{k}} \sum_{\vec{q}} \sum_{\sigma_1} \sum_{\sigma_2} n_{\sigma_1}(\vec{k}+\vec{q}) n_{\sigma_2}(\vec{k}) Y_{20}(-\hat{q}) \overline{N}_{12}^{(0)}(\sigma_1, \sigma_2) \tag{16}$$

where

$$h_0 = \frac{4\pi}{3}\frac{\sqrt{16\pi}}{\sqrt{5}} ; \overline{N}_{12}^{(0)}(\sigma_1,\sigma_2) = \frac{1}{4}\delta_{\sigma_1,\sigma_2} - \frac{1}{4}(\delta_{\sigma_1,\downarrow}\delta_{\sigma_2,\uparrow} + \delta_{\sigma_1,\uparrow}\delta_{\sigma_2,\downarrow}) \quad (17)$$

In terms of the Legendre polynomial, one has

$$Y_{20}(\hat{q}) = Y_{20}(-\hat{q}) = (5/4\pi)^{1/2}P_2(\cos\theta_{\hat{q}}) = (5/4\pi)^{1/2}(1/2)(3\cos\theta_{\hat{q}}^2 - 1) \quad (18)$$

As in the case of the familiar electron gas problem with uniform positive ionic background, there is no contribution due to the first (direct) term of eq. (9), because $V_{12}(\vec{q}=0) = 0$.

Note that in any variational paramagnetic state, with $n_\uparrow(\vec{k}) = n_\downarrow(\vec{k})$, the exchange contribution goes to zero because of summations over spins $\sigma_1$ and $\sigma_2$, due to the particular form of the spin matrix elements given in eq.(17). There is a contribution only from the kinetic energy part, and the best variational paramagnetic state is then nothing but the non interacting uniform paramagnetic state with energy $E_0$.

The exchange term of eq.(16) gives the maximum negative contribution when $\sigma_1 = \sigma_2$, and the direction of the momentum transfer is such that $\cos\theta_{\hat{q}}^2 < 1/3$, i.e., close to the spin quantization direction $\hat{z}$. Note that the summations over $\vec{k}$ and $\vec{q}$ give zero for the exchange contribution if the occupation function is spherical. To proceed further, let us assume that all spins are parallel (say, in the up-direction), with $n_\downarrow(\vec{k}) = 0$, but $n_\uparrow(\vec{k})$ depends on both the magnitude and the direction of $\vec{k}$. For example, one may consider the form

$$n_\uparrow(\vec{k}) = \Theta(k_{F\uparrow}^2 - k^2 + k^2(\delta_1 Y_{10}(\hat{k}) + \delta_2 Y_{20}(\hat{k}) + ...)) \; ; \sum_{\vec{k}} n_\uparrow(\vec{k}) = N \quad (19)$$

in which $\delta_1, \delta_2,...,$ lead to deformation from the spherical surface. Before proceeding further, it has to be noted that following the usual procedure [3] a simple calculation of the proper self-energy $\hbar\Sigma_\uparrow^*(\vec{k})$ can be easily done to the lowest order in the dipole-dipole interaction, for a system of particles with only up-spins, starting with the unperturbed Green's function of the non interacting gas corresponding to the occupation function $n_{0\sigma}(k) = \Theta(k_{F\uparrow} - k)\delta_{\sigma,\uparrow}$. Such a calculation shows that the resulting proper self-energy is proportional to $(-)Y_{20}(\hat{k})$ with a positive proportionality factor which is a function of the magnitude $k$. This implies that in the polar direction, the single particle energy $(\hbar^2/2m)k^2 + \hbar\Sigma_{\uparrow}^*(\vec{k})$ is lower compared to its value in the equatorial $k_x - k_y$ plane for the same value of the magnitude of the wave vector. Note, however, that the contribution to the total energy from this proper self-energy goes to zero, as already observed, since it requires the integration over $\vec{k}$ of the product of the self-energy and the unperturbed spherical function $n_{o\uparrow}(k)$. Thus, there is no alternative but to deform the occupation function from the spherical shape to get any non vanishing contribution to $E_{exch}$. However, the simple self-energy calculation does suggest that it is more natural to consider the deformation of the surface bounding the occupied states to be of the quadrupolar type corresponding to a prolate spheroid. We consider this case first.

*2.1.1 Quadrupolar Deformation*

Let the occupied region of $\vec{k}$ is a prolate spheroid (a symmetrical egg) pointed towards the z-direction. The surface bounding the occupied region in the $\vec{k}$-space is then given by the equation

$$\frac{(k_x^2 + k_y^2)}{k_{Fx}^2} + \frac{k_z^2}{k_{Fz}^2} = 1 \quad , k_{Fz} > k_{Fx} \tag{20}$$

so that

$$n_{\uparrow}(\vec{k}) = \Theta(1 - \frac{(k_x^2 + k_y^2)}{k_{Fx}^2} - \frac{k_z^2}{k_{Fz}^2}) = \Theta(k_{F\uparrow}^2 - k^2(1 - \beta_2 P_2(\cos\theta_{\hat{k}}))) \tag{21}$$

in this state (to be called the JM ferromagnetic state), where,

$$k_{F\uparrow}^2 = \frac{3k_{Fx}^2 k_{Fz}^2}{(2k_{Fz}^2 + k_{Fx}^2)} \quad ; \beta_2 \equiv (5/4\pi)^{1/2}\delta_2 = \frac{2(k_{Fz}^2 - k_{Fx}^2)}{(2k_{Fz}^2 + k_{Fx}^2)} \tag{22a}$$

$$k_{Fx}^2 = k_{F\uparrow}^2/(1 + \beta_2/2) \quad ; k_{Fz}^2 = k_{F\uparrow}^2/(1 - \beta_2) \tag{22b}$$

and where the volume of the k-space spheroid is given by

$$V_{spheroid} = (4\pi/3)k_{Fx}^2 k_{Fz} = (4\pi/3)k_{F\uparrow}^3 / [(1+\beta_2/2)(1-\beta_2)^{1/2}] \quad (22c)$$

Using the fact that for only up- spins present in the system,

$$\sum_{\sigma_1}\sum_{\sigma_2} \bar{N}_{12}^{(0)}(\sigma_1, \sigma_2) = 1/4 \quad (23)$$

$E_{exch}$ for different types of deformation can be calculated from the expression

$$E_{exch} = -(\mu^2/2)h_0 \frac{1}{4V_0} \sum_{\vec{k}}\sum_{\vec{q}} n_\uparrow(\vec{k}+\vec{q})n_\uparrow(\vec{k})Y_{20}(-\hat{q}) \quad (24)$$

If we use the form (21) for $n_\uparrow(\vec{k})$ to calculate the exchange energy $E_{exch}^{(Q)}$ in the JM state, with positive deformation parameter $\beta_2$, for occupied states one has to restrict its value such that $0 \le \beta_2 \le 1$. For $\beta_2 = 0$, the surface becomes spherical whereas for $\beta_2 = 1$, one finds $k_{Fz} \to 0$, a singular behavior, but the volume of the spheroid is still finite with $k_{Fx}^2 k_{Fz}/6\pi^2 = n = N/V_0$. We propose to consider the value of the deformation parameter such that $0 < \beta_2 < 1$. The form (21) leads to the expression for the number density as

$$\frac{N}{V_0} = \frac{1}{V_0}\sum_{\vec{k}} n_\uparrow(\vec{k}) = \frac{V_{spheroid}}{8\pi^3} = \frac{k_{F\uparrow}^3}{6\pi^2}\left[\frac{1}{(1+\beta_2/2)(1-\beta_2)^{1/2}}\right] \to \frac{k_{F\uparrow}^3}{6\pi^2}(1+\frac{3}{8}\beta_2^2 + ..) \quad (25)$$

which has no correction to the linear order in $\beta_2$ for small deformations, compared the spherical case ($\beta_2 = 0$) with the non interacting gas ferromagnetic state Fermi wavevector $k_{F\uparrow} = 2^{1/3}k_{F0}$, where $k_{F0}$ is the Fermi wave vector for the non interacting particles in the uniform paramagnetic state. Here, $N/V_0 = k_{F0}^3/3\pi^2$. Substituting the form(21) in eq. (15),the kinetic energy can be calculated exactly. One finds,

$$E_{kin}^{(Q)}(\text{JM}) = N\frac{\hbar^2 k_{F\uparrow}^2}{2m}(3/5)\frac{(1-\beta_2/2)}{(1+\beta_2/2)(1-\beta_2)} \to E_0\{2^{2/3}[1+(1/4)\beta_2^2]\} \quad (26)$$

where

$$E_0 = (3/5)\frac{\hbar^2 k_{F0}^2}{2m} = (2.21/r_0^2)\frac{\hbar^2}{2m} \quad (27)$$

is the energy of the best variational paramagnetic state with $n_\sigma(k) = \Theta(k_{F0} - k)$, which is just the ground state energy of non interacting spin-half Fermi gas. For a general value of $\beta_2$ between 0 and 1, it is not easy to obtain the exchange energy contribution analytically. However, for small deformations one can expand the expression (21) for the occupation function in powers of $\beta_2$. As we have seen already, to the order linear in $\beta_2$ there are no corrections to the number density and to the kinetic energy. But, in the lowest order the exchange energy is linear in $\beta_2$, and a straight forward calculation leads to

$$E_{exch}^{(Q)}(JM) = -(\mu^2/r_0^3)\beta_2(9/100)J^{(Q)} \; ; \; J^{(Q)} = 5/6 \qquad (28)$$

The total variational energy per particle to the lowest order in $\beta_2$ then becomes

$$\frac{E^{(Q)}}{N}(JM) = \frac{E_{kin}^{(Q)}}{N} + \frac{E_{exch}^{N}}{N} = \frac{E_0}{N}\{2^{2/3} - (3/40)\beta_2 \frac{r_m}{(2.21)r_0}\} \qquad (29)$$

For different allowed positive values of the quadrupolar deformation parameter $\beta_2$, Fig.1 shows a plot of $E^{(Q)}/E_0$, as a function of the dimensionless density parameter $r_{sm} = (2.21)r_0/r_m$. The present result for the so called JM state is different quantitatively than our earlier result (eq. 3.28 of Ref.1), where we had taken the form of the single particle density operator $\rho^{(1)}$ and $n_\uparrow(\vec{k})$ as given by eqs (4) and (11). However, qualitatively results are similar. In fact, if we identify the earlier deformation parameter $\beta$ of Eq.(11) with $\beta_2$ used in the determinantal calculation here, for the same density parameter $r_{sm}$ our present calculation gives lower variational energy (see Fig.1, where both results are plotted). Since the present calculation of variational HF energy is an upper bound to the ground state energy, it turns out that the result in Ref.1 was also an upper bound for the particular choice of $\rho^{(1)}$, but is not as good as the present HF case! In the present case, the total energy becomes less than $E_0$ of the uniform variational paramagnetic state for

$$r_{sm} < \frac{3}{40}[\beta_2/(2^{2/3}-1)], \text{ i.e., } r_0 < \frac{3}{40}\frac{\beta_2 r_m}{(2^{2/3}-1)(2.21)} \qquad (30)$$

When we increase the density further, the total energy becomes negative for

$$r_0(critical) \leq \frac{3}{40} \frac{\beta_2 r_m}{(2^{2/3})(2.21)} \tag{31}$$

and, as before, eventually the system will collapse to an infinite density state.

*2.1.2 Dipolar deformation*

Although, the quadrupolar deformation was the most natural choice, it is instructive to consider also a purely dipolar deformation of the surface bounding the occupied spin-up only states in the $\vec{k}$-space. In this case, let us assume

$$n_\uparrow(\vec{k}) = \Theta(k_{F\uparrow} - k\sqrt{1 - \delta_1 Y_{10}(\hat{k})}) = \Theta(k_{F\uparrow}^2 - k^2 + k^2 \beta_1 P_1(\cos\theta_{\hat{k}})) \tag{32}$$

where

$$\beta_1 = \sqrt{3/4\pi}\, \delta_1, P_1(\cos\theta_{\hat{k}}) = \cos\theta_{\hat{k}} = \sqrt{4\pi/3}\, Y_{10}(\hat{k}). \tag{33}$$

When the above form of $n_\uparrow(\vec{k})$ is substituted in eqs (3c), (15) and (24) to calculate the number density, the kinetic energy and the exchange energy respectively, we again find that whereas the number density and the kinetic energy contribution can be calculated exactly as a function of deformation parameter $\beta_1$, the magnitude of which can take values between 0 and 1, it is not easy to obtain the exact exchange energy contribution analytically. Thus, for calculating the exchange energy we expand the occupation function given by eq. (32) in powers of $\beta_1$, for small deformations. As expected, in the lowest order this contribution is of the order $\beta_1^2$. The final results of our calculations are as follows,

$$\frac{N}{V_0} \to \frac{k_{F\uparrow}^3(dip.)}{6\pi^2}[1 + (5/8)\beta_1^2]\; ; k_{F\uparrow}^3(dip.) \to k_{F0}^3\, 2[1 + (5/8)\beta_1^2]^{-1} \tag{34}$$

$$\frac{E_{kin}^{(dip.)}}{N} \to \frac{E_0}{N} 2^{2/3}[1 + (5/12)\beta_1^2] \tag{35}$$

$$\frac{E_{exch}^{(dip.)}}{N} \to -\frac{\mu^2}{r_0^3}\frac{\beta_1^2}{20} \tag{36}$$

The total energy correct to the order $\beta_1^2$ is then given by

$$\frac{E^{(dip.)}}{N} = \frac{E_0}{N}[2^{2/3}(1 + (5/12)\beta_1^2) - \frac{\beta_1^2}{20(2.21)}\frac{r_m}{r_0}] \tag{37}$$

The above energy is lower than the uniform paramagnetic state energy $E_0/N$, for

$$r_0 < \frac{20\beta_1^2 r_m}{2^{2/3}(2.21)[1+(5/12)\beta_1^2]}. \tag{38}$$

As the density is increased further, the total energy with dipolar deformation also becomes negative, and eventually collapses to the infinite density state. However, note that the comparison of expressions (29) and (37) shows that at the same density $E^{(dip.)}/N$ is higher than $E^{(Q)}/N$ of the JM state, even if one takes $|\beta_1|$ high enough so that $\beta_1^2 \approx \beta_2$. In Fig.2, we have plotted both these determinantal HF results for an easy comparison. For small deformations, the result that the JM state involving quadrupolar deformation gives a lower energy is of course valid exactly, and one can not really give too much importance to the fact that a comparison of these approximate expressions even for higher allowed values (close to 1) for the respective deformation parameters $\beta_1^2, \beta_2$ gives lower energy for the quadrupolar case. But, it seems very likely that the quadrupolar deformation of the surface bounding the occupied spin-up states in the JM state gives a better upper bound to the ground state energy.

## 3. Variational HF ground state energy of the magnetic dipole system in presence of a hard core potential

Now, we consider the situation in which the chargeless fermion system with magnetic dipole interaction has also a short range repulsive hard core interaction between the particles. This is to explore whether the inclusion of this hard core interaction can stop the high density collapse of the JM ferromagnetic state with spheroidal shape in the $\vec{k}$ space of the occupied spin-up states. The most general form of the velocity-independent interaction between spin-half particles can be written in the form[3]

$$V_{12}(\vec{r}) = V_c(\vec{r}) + V_s(\vec{r})\vec{s}_1.\vec{s}_2 + V_T(\vec{r})[\vec{s}_1.\vec{s}_2 - 3\vec{s}_1.\hat{r}\vec{s}_2.\hat{r}] \tag{39}$$

The last term (tensor interaction) has the form of the magnetic dipole-dipole interaction in our problem, with $V_T = \mu^2/r^3$. We had no other interaction until now. We will now add a very short range (finite) highly repulsive interaction for calculating its expectation value in the following chosen determinantal states :

(a): The uniform paramagnetic state ( PARA): $n_\sigma(\vec{k}) = n_0(k) = \Theta(k_{F0} - k)$  (40)

(b): The fully polarized ferromagnetic state with quadrupolar deformation ( JM state):

$$n_\sigma(\vec{k}) = n_\uparrow(\vec{k})\delta_{\sigma,\uparrow} \; ; n_\uparrow(\vec{k}) \equiv \Theta(k_{F\uparrow}^2 - k^2 + k^2\beta_2 P_2(\cos\theta_{\vec{k}})) \quad (41)$$

We can choose the central short range repulsive part, of the form as in the first or in the second term of eq. (39). In the PARA state, the second term does not contribute to the direct part in its expectation value, because of the spin summations. The exchange part has a factor of ½ due to spin summations. In the JM state, except for an additional factor of ¼ due to the spin part, the contribution will be similar to the first term. Thus it is enough to consider the form of the only first term in eq. (39) for the repulsive short range part, to get all the results we want. Let,

$$V_c(\vec{r}) = U_0 \text{ ( large \& positive) for } r \leq a \; , \; V_c(\vec{r}) = 0, r > a \quad (42)$$

The range $a$ is assumed to be small but finite. The Fourier transform in the $\vec{q}$ – space of the above interaction is

$$V_c(\vec{q}) = \int d^3 r e^{-i\vec{q}\cdot\vec{r}} V_c(\vec{r}) = V_c(q=0)[3(\sin qa - qa\cos qa)/q^3 a^3] \quad (43)$$

where,

$$V_c(q=0) \equiv V_c(0) = (4\pi/3)U_0 a^3 \; ; \; U_0 a^3 \equiv (\hbar^2/2m) 6 a_0 \quad (44)$$

Note that there are two parameters in the potential, the strength $U_0$ and the range a. Sometimes, one replaces the coupling constant $U_0 a^3$ by the actual low-energy scattering length $a_0$ [3], as indicated in eq. (44).

In general, the expectation value of the hard core interaction (43) in any HF state is

$$E_c = \frac{V_0}{2} \int (d^3 k_1/8\pi^3) \int (d^3 k_2/8\pi^3)$$
$$\times \sum_{\sigma_1}\sum_{\sigma_2} n_{\sigma_1}(\vec{k}_1) n_{\sigma_2}(\vec{k}_2)[V_c(0) - \delta_{\sigma_1,\sigma_2} V_c(\vec{k}_1 - \vec{k}_2 \equiv \vec{q})], \; V_0 \equiv volume \quad (45)$$

(a) In the PARA state, this gives :

$$E_c(para) = \frac{V_0}{2} \times 2 \int (d^3 k_1/8\pi^3) \int (d^3 k_2/8\pi^3)[2V_c(0) - V_c(q)] \quad (46)$$

This leads to the usual result valid for any potential not depending on the direction of $\vec{q}$:

$$E_c(para)/N = \frac{1}{2} \times \frac{1}{2\pi^3} \int d^3 q (1 - \frac{3}{2} x + \frac{1}{2} x^3) \Theta(1-x) [2V_c(0) - V_c(q)] \; ; x \equiv \frac{q}{2k_{F0}} \quad (47)$$

(b) In the JM state, we have the general result

$$E_c(JM) = \frac{V_0}{2(2\pi)^6} \int d^3k \int d^3q \; n_\uparrow(\vec{k}) n_\uparrow(\vec{k}+\vec{q}) [V_c(0) - V_c(q)] \quad (48)$$

To the leading orders in $k_{F0} a$, i.e. $a/r_0 \ll 1$, the expression (46) for the paramagnetic state goes as $(a/r_0)^3$, whereas the expression (47) for the JM state goes as $(a/r_0)^5$, both being positive. Since the magnetic dipole contribution in the JM state, as calculated in section 2 varies as $1/r_0^3$ with a negative sign, at high densities it is the positive hard core contribution which will dominate the total energy, as long as the range $a \ll r_0$. However, this condition is unnecessary for having a stable JM state, as we will show below.

More generally, to the linear order in the quadrupolar deformation parameter $\beta_2$ of the JM state, there is no need to recalculate the kinetic energy term and the dipolar exchange term. They remain the same as in section 2. One has to recalculate only the repulsive core contribution without any expansion in the parameter $a/r_0$. We find

$$E_c(para)/E_0 = \frac{2mU_0 a^2}{\hbar^2} (\frac{19.2}{54\pi})(\frac{a}{r_0})[1 + T(\alpha_0)] \; ; \alpha_0 \equiv 2k_{F0} a = 2(\frac{1.92a}{r_0}) \; ;$$

$$E_0 \equiv (\hbar^2/2m)(2.21/r_0^2) \; ; 1.92 \cong (9\pi/4)^{1/3} \quad (49)$$

where the function T( s) is defined by,

$$T(s) \equiv 1 - \frac{72}{s^3}[(Si(s) - \sin s) - \frac{3}{2s}(2 - 2\cos s - s \sin s)$$
$$+ \frac{1}{2s^3}(8s \sin s + 8\cos s - 8 - 4s^2 \cos s - s^3 \sin s)] \quad (50)$$

and where the sine integral function

$$Si(s) = \int_0^s dx (\sin x / x) = s - \frac{s^3}{3!3} + \frac{s^5}{5!5} - \frac{s^7}{7!7} + .. \quad (51)$$

Similarly, one finds

$$E_c(JM)/E_0 = \frac{2mU_0 a^2}{\hbar^2}(\tfrac{19.2}{27\pi})(\tfrac{a}{r_0})T(\alpha_\uparrow) \ ; \alpha_\uparrow \equiv 2k_{F\uparrow}a = 2^{4/3}(\tfrac{1.92a}{r_0}) \quad (52)$$

Thus in the presence of the hard core interaction, the expression (29) giving the ratio of the total energy E and $E_0$ in the JM state is replaced by

$$\frac{E}{E_0}(JM) = \{2^{2/3} + \frac{(2m/\hbar^2)U_0 a^3}{r_0}(\tfrac{19.2}{27\pi})T(\alpha_\uparrow) - (3/40)\beta_2 \frac{(2m/\hbar^2)\mu^2}{(2.21)r_0}\} \quad (53)$$

where, as defined before, $r_m \equiv 2m\mu^2/\hbar^2$ ; $r_{sm} = (2.21)r_0/r_m$.

From the definition of the function T (s) given by eqs (50) and (51), it is easy to see that it is a smooth function, with $T(s) \to 1$, as $s \to \infty$ and $T(s) \to 3s^2/100$, for $s \ll 1$. With the definition given in eq. (52) for $\alpha_\uparrow \sim 1/r_0$, the expression (53) immediately shows us that for the ratio $U_0 a^3/\mu^2$ of the coupling constants not extremely small compared to 1, the total energy becomes positive as $r_0 \to 0$, and there is no longer the high density collapse of section 2, as explicitly shown in Fig.3. The JM ferromagnetic state is now a possible stable equilibrium ground state of the system.

## 4. Concluding remarks

Our interest in the nature of the ground state of chargeless fermions with a finite magnetic moment initially arose because of the suggestion by Yajnik [4] that the state of the universal relic background neutrinos might be a ferromagnetic state with domain walls, made in the context of big-bang cosmology. However, we find that in the non relativistic case, at T = 0 K, the density required for the ferromagnetic transition, with a spheroidal occupation function, is too high for satisfying the condition $r_0 < r_m = 2\mu^2 m/\hbar^2$. For chargeless fermions with an atomic mass of $10^4$ to $10^5$ times the electron mass and magnetic moment $\mu$ of the order of the Bohr magneton, $r_m \approx 10^{-9} - 10^{-8}$ cm. For neutrinos with a mass in the range of 0.01eV and a very tiny magnetic moment [5], $r_m$ is extremely small. May be, in the cosmological context it is still interesting to investigate this problem in the relativistic regime and at high temperatures, conditions relevant to the times very close to the big bang. Also, in general it will be important to find numerically what is the optimum value of the only variational parameter $\beta_2$ which gives the lowest upper bound for the ground state energy, and what is the corresponding equilibrium density for the JM ferromagnetic state for given hard core and magnetic dipole interactions. The positive kinetic energy per particle tends to infinity as the parameter $\beta_2$ tends to 1. In that case, to minimize the kinetic energy one may have to localize the particles on a lattice [6]. We find that the problem of spin-dependent non-central interaction in itself may be of great intrinsic interest for physics of many-particle systems. In nuclear physics, the non-central tensor interaction [3] which explains the finite quadrupole moment for deuteron has similar spin structure, but that is supposed to be only a small perturbation to the main short-range central force. The dipole-dipole interaction is also considered as a perturbation in most condensed matter physics problems. In our case, as we have shown in this paper, the magnetic dipole-dipole interaction is the main player to lead to a stable equilibrium ferromagnetic JM state in the presence of a very short range repulsive finite hard core interaction of sufficient strength.

**Acknowledgment**

One (SSJ) of us would like to thank the DAE, Government of India, for the Raja Ramanna Fellowship awarded to him to work at IIT Bombay, and would like to acknowledge support from the condensed matter theory group of the Department of Physics and Astronomy at Michigan State University. We thank Khang Hoang for helping us with the figures.

**Figure captions**

**Figure 1**. Plot of the ratio of the total variational energy $E$ in the ferromagnetic state and the variational energy $E_0$ in the uniform paramagnetic state, as a function of the density parameter $r_{sm} = (2.21)r_0/r_m$, in the absence of any hard core interaction. Solid curves represent the determinantal HF calculations in the JM ferromagnetic state for different parameters $\beta_2$ representing quadrupolar deformation of the surface bounding the occupied states $n_\uparrow(\vec{k})=1$, and dotted curves represent our earlier calculation [1] using a variational single particle density matrix $\rho^{(1)}$ with quadrupolar deformation parameter $\beta$.

**Figure 2**. Comparison of the variational energy $E$ in the ferromagnetic state calculated using determinantal HF wave functions with purely quadrupolar and purely dipolar deformations of the occupation function, as a function of the density parameter $r_{sm} = (2.21)r_0/r_m$, in the absence of any hard core interaction. The solid curves represent the ratio $E^{(Q)}/E_0$ for different quadrupolar deformation parameters $\beta_2$ in the JM state, whereas the dash-dot curves represent the ratio $E^{(dip.)}/E_0$ for the square of different dipolar parameters $\beta_1$.

**Figure 3.** Plot of the total variational HF energy $E$ in the ferromagnetic JM state, in units of the enrgy $E_0$ of the corresponding paramagnetic noninteracting gas, in the presence of the magnetic dipole interaction as well as a repulsive short range hard core interaction of range $a$ and strength $U_0$, as a function of $r_0/r_m$. Plots are for two values of the ratio of the two coupling constants $\Gamma_{cm} \equiv U_0 a^3/\mu^2$, in which the label $p \equiv a/r_m$ is the range $a$ in units of the magnetic length $r_m \equiv (2m\mu^2/\hbar^2)$.

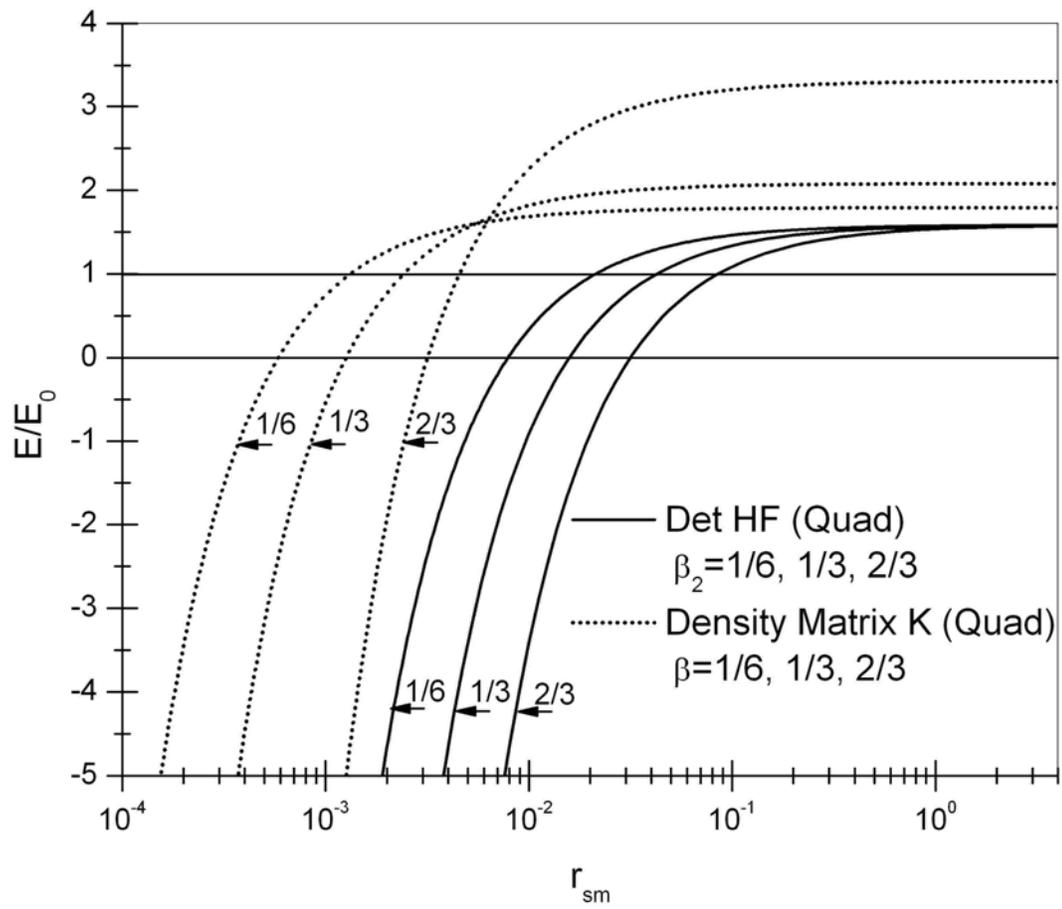

Figure 1

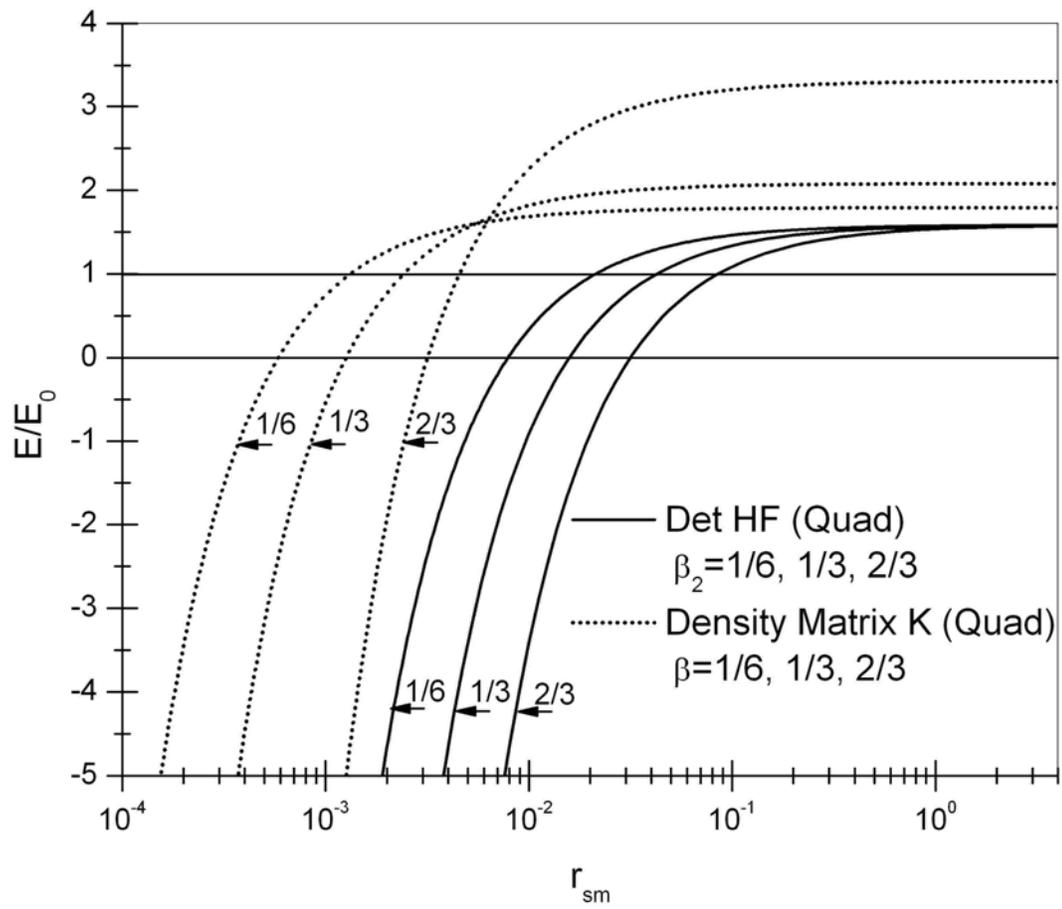

Figure 1

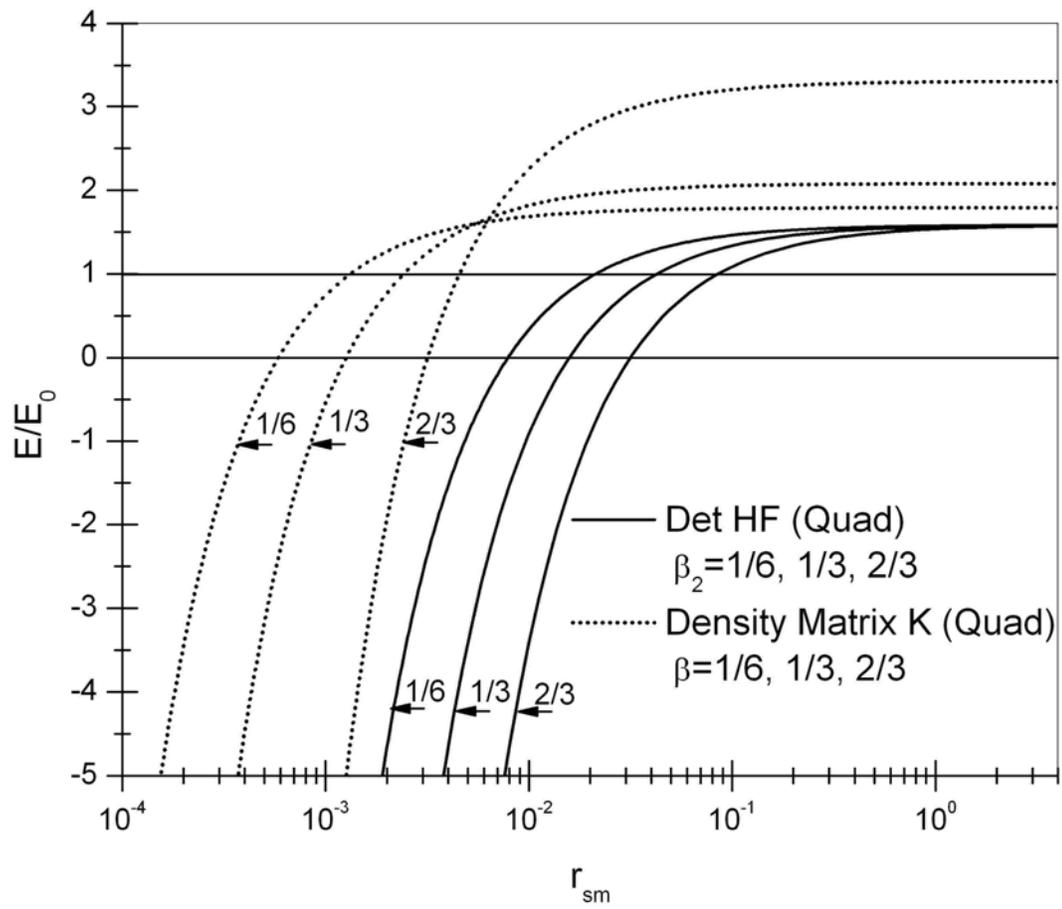

Figure 1

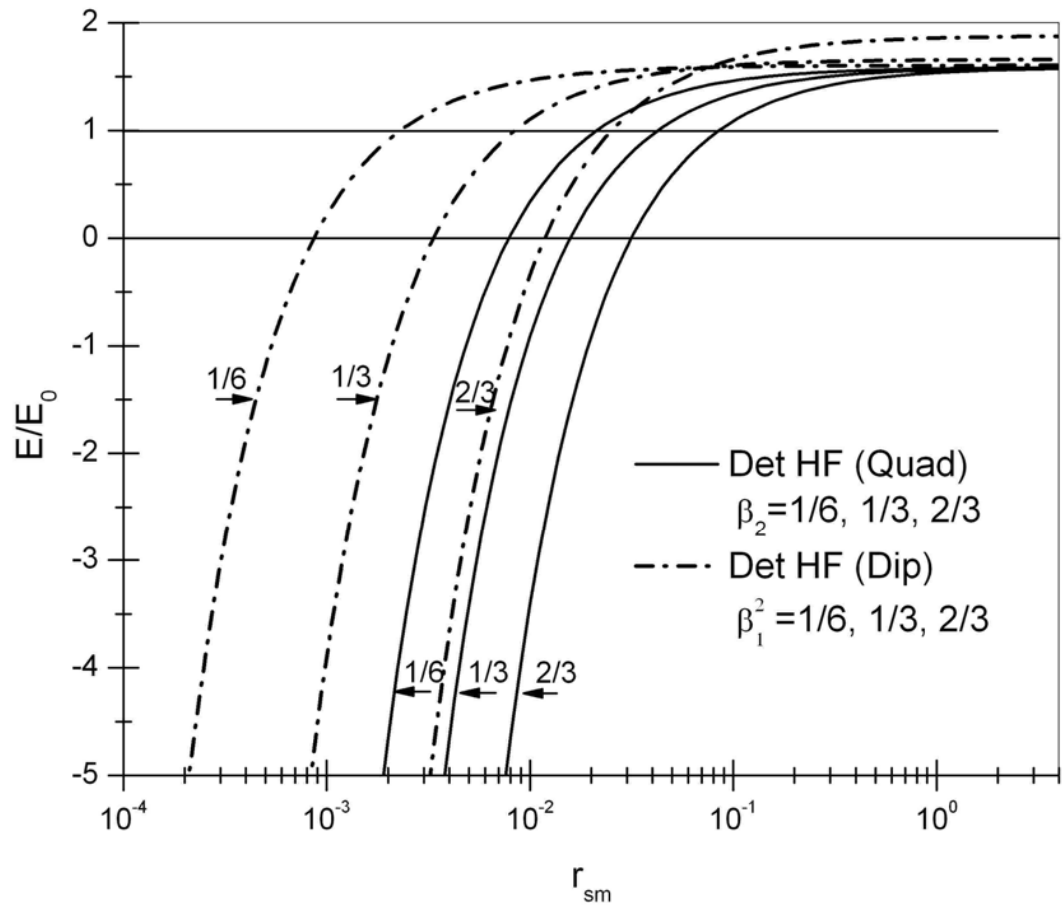

Figure 2

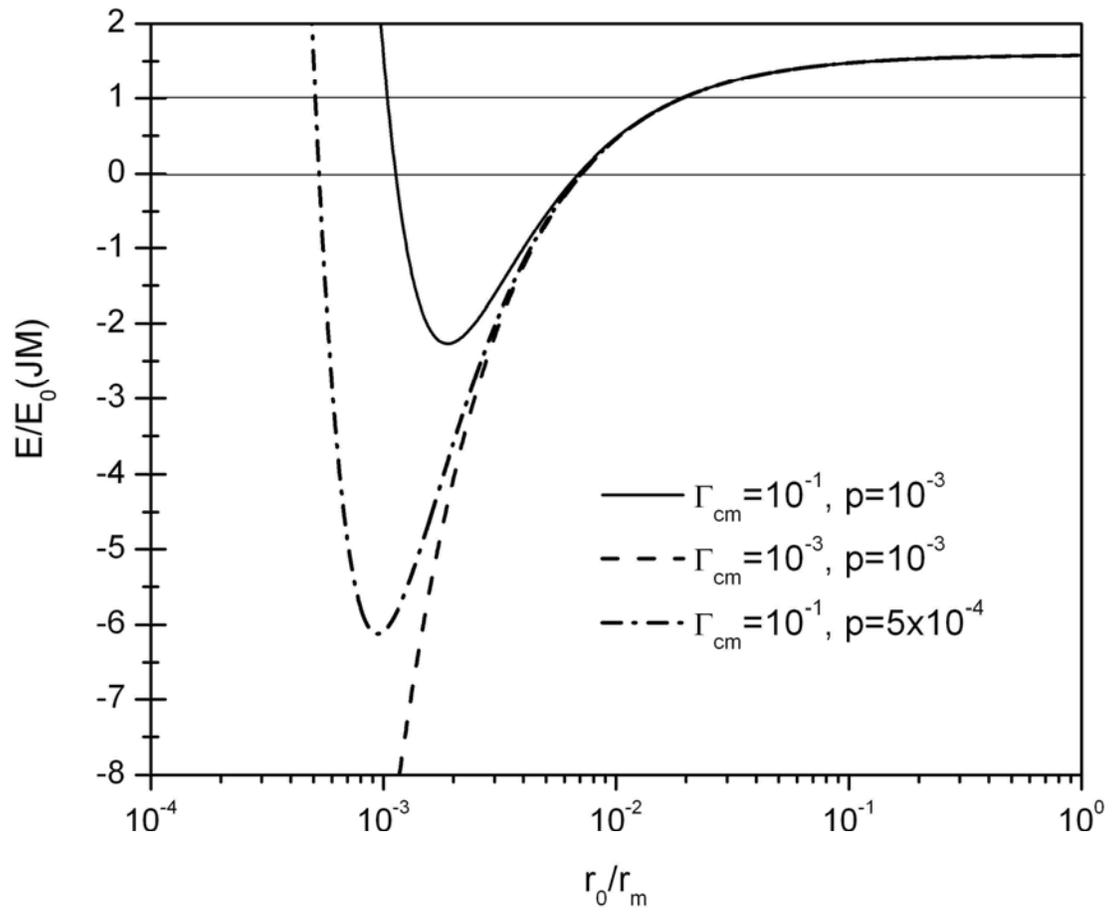

Figure 3